\documentclass[twocolumn]{jpsj3}
\usepackage{color}
\usepackage{txfonts}
\usepackage{bm}

\title{Multi-orbital Fulde-Ferrell-Larkin-Ovchinnikov State in SrTiO$_3$ Heterostructures}

\author{ \name{Yasuharu Nakamura}$^1$ and \name{Youichi Yanase}$^{1,2}$\thanks{E-mail address: yanase@phys.sc.niigata-u.ac.jp}}

\inst{$^1$Graduate School of Science and Technology, Niigata University, Niigata 950-2181, Japan \\
$^2$Department of Physics, Niigata University, Niigata 950-2181, Japan} 

\recdate{August 11, 2014}

\abst{
We study exotic superconducting states in SrTiO$_3$ heterostructures on the basis of the three-orbital model, 
which reproduces the band structure of two-dimensional electron gases. 
We show various Fulde-Ferrell-Larkin-Ovchinnikov (FFLO) states induced by  
the broken inversion symmetry and orbital degree of freedom. 
In particular, a novel orbital-dependent FFLO state is stabilized in a magnetic field along the [110]-axis. 
The field angle dependence of the FFLO state is clarified on the basis of the spin and orbital texture in the momentum space. 
It is shown that the orbital degree of freedom in Cooper pairs gives rise to in-plane anisotropy of 
the critical magnetic field. The carrier density dependence of the superconducting state is also discussed. 
}

\begin{document}
\maketitle

\newcommand{\us}{\uparrow}
\newcommand{\ds}{\downarrow}
\newcommand{\g}{\mbox{\boldmath$g$}}
\renewcommand{\k}{\mbox{\boldmath$k$}}
\newcommand{\LL}{\mbox{\boldmath$L$}}
\renewcommand{\SS}{\mbox{\boldmath$S$}}
\newcommand{\etal}{{\it et al.}}
\newcommand{\q}{\bm{q}}
\newcommand{\rr}{\bm{r}}
\newcommand{\HH}{\bm{H}}

\section{Introduction}

Since the discovery of two-dimensional conducting electron gases at the interface between 
two band insulator perovskite oxides SrTiO$_3$/LaAlO$_3$,~\cite{Ohtomo}
quantum phases in SrTiO$_3$ heterostructures have been explored extensively.  
Superconductivity has been found not only in the SrTiO$_3$/LaAlO$_3$ interface~\cite{Reyren} 
but also on the SrTiO$_3$ surface induced by an electric double-layer transistor (EDLT).~\cite{Ueno} 
Superconductivity also occurs in the SrTiO$_3$/LaTiO$_3$ interface~\cite{Biscaras} 
and $\delta$-doped SrTiO$_3$.~\cite{Kozuka} 
Interestingly, these superconducting states are artificially tuned by 
the gate voltage~\cite{Ueno,Ueno_review,Caviglia,Bell,Shalom,Caviglia-2}. 
The EDLT technique also realized electric-field-induced superconductivity in KTaO$_3$, MoS$_2$, ZrNCl, 
and La$_{1-x}$Sr$_x$CuO$_4$.~\cite{Ueno_review} 
In this paper, we propose exotic superconducting states that appear in SrTiO$_3$ heterostructures.

 Superconductivity induced by the condensate of Cooper pairs having a finite center-of-mass momentum was proposed by 
Fulde and Ferrell~\cite{FF} and by Larkin and Ovchinnikov~\cite{LO} five decades ago,~\cite{FFLO_review} 
although the standard BCS theory assumes a zero center-of-mass momentum in Cooper pairs~\cite{BCS}. 
 Experimental searches for the Fulde-Ferrell-Larkin-Ovchinnikov (FFLO) state found 
the heavy-fermion superconductor CeCoIn$_5$,~\cite{PhysRevLett.91.187004,Kenzelmann} 
some organic superconductors,~\cite{PhysRevLett.97.157001,PhysRevLett.99.187002,PhysRevB.83.064506,Mitrovic} 
and iron-based superconductors~\cite{Terashima,Burger,Zocco} to be promising candidates. 
The FFLO states in population-imbalanced cold fermion gases~\cite{Nature.467.567} 
and in nuclear matter~\cite{RevModPhys.76.263} have also attracted interest.

 For the FFLO state rather than the BCS state to be stabilized, the spin polarization must be caused by 
something that breaks the time-reversal symmetry, such as an applied magnetic field or 
proximity to a ferromagnet.~\cite{FF,LO,FFLO_review} 
A magnetic field is most easily achieved, but it simultaneously leads to the orbital depairing effect 
and often destabilizes the FFLO state.~\cite{Gruengerg,Adachi} In the above candidate materials for the FFLO superconductor, 
the heavy and anisotropic effective mass of quasiparticles suppresses the orbital depairing effect and allows the FFLO state
to be stabilized. 
On the other hand, the orbital depairing effect is completely suppressed 
when a magnetic field is applied parallel to the conducting plane of two-dimensional electron gases. 
Therefore, SrTiO$_3$ heterostructures are a valuable platform for realizing the FFLO state.

The artificial tuning of the superconducting state using the gate 
voltage~\cite{Ueno,Ueno_review,Caviglia,Bell,Shalom,Caviglia-2} may enable a novel FFLO state to emerge. 
For instance, the lack of space inversion symmetry on the surface/interface allows an FFLO state 
beyond the paradigm of Fulde and Ferrell and of Larkin and Ovchinnikov. 
For clarity, we adopt the following classification of the FFLO state. 
The Fulde-Ferrell (FF) state is the single-$Q$ condensate represented by the order parameter $\Delta(\rr) = \Delta \, e^{i \q \rr}$ 
in real space,~\cite{FF} whereas the Larkin-Ovchinnikov (LO) state is the double-$Q$ state where 
$\Delta(\rr) = \Delta \left(e^{i \q \rr} + e^{-i \q \rr}\right)/2 = \Delta \cos(\q \rr) $.~\cite{LO} 
While the phase of the order parameter shows spatial modulation in the FF state, the order parameter in the LO state 
acquires amplitude modulation. 
Furthermore, it has been shown that the triple-$Q$, quadruple-$Q$, and sextuple-$Q$ states are stabilized  
when the Fermi surface has cylindrical symmetry.~\cite{Shimahara_multiple-Q} 

 Recent theoretical studies on superconductivity lacking inversion symmetry, 
which is called noncentrosymmetric superconductivity,~\cite{NCSC} elucidated a ``helical superconducting state'' similar to the 
FF state.~\cite{SovPhysJETP.68.1244,JETPLett.78.637,PhysRevB.70.104521,PhysRevLett.94.137002,PhysRevB.75.064511,PhysRevB.76.014522} 
Owing to the antisymmetric spin-orbit coupling appearing in noncentrosymmetric crystals,~\cite{NCSC}
the helical state is stabilized in the low-magnetic-field region above $H_{\rm c1}$, 
in contrast to the FFLO state which requires a high magnetic field close to the Pauli limit to be stabilized in 
centrosymmetric crystals. 
It has also been shown that an intermediate state between the FF and LO states, 
which we call the ``complex stripe (CS) state'', emerges from the helical state.~\cite{PhysRevB.75.064511}
A helical state robust against spin polarization has been proposed for the SrTiO$_3$/LaAlO$_3$ interface~\cite{Michaeli} 
in order to elucidate the coexistence of superconductivity and ferromagnetic order.~\cite{Brinkman,Dikin,Li,Bert,Ariando} 

Although these previous theories are based on single-band and single-orbital models, 
it has been shown that the orbital degree of freedom in $t_{2g}$ electrons plays an important role 
in the magnetic response of superconducting SrTiO$_3$ heterostructures.~\cite{Nakamura_STO} 
Thus, a further novel FFLO state may be induced by the cooperation between the broken inversion symmetry 
and the orbital degree of freedom. 
Although the FFLO superconductivity in multi-band models has been investigated 
in recent studies~\cite{Gurevich,Ptok,Mizushima} triggered by the observation of a paramagnetic depairing effect 
in iron-based superconductors,~\cite{Terashima,Burger,Zocco} 
the orbital degree of freedom has not been taken into account. 
When we assume a pairing interaction on the band basis as in the literature,~\cite{Gurevich,Ptok,Mizushima} 
the momentum dependence of the orbital wave function in the band is neglected. 
On the other hand, it is naturally taken into account in the multi-orbital model that we adopt in this paper. 
Furthermore, the Rashba spin-orbit coupling, which plays an essential role in this study, is appropriately derived 
in the multi-orbital model,~\cite{Yanase_NCSC} although it is difficult to do so in multi-band models. 
Thus, the multi-orbital model is a reasonable starting point in this study. 
We show that the multi-orbital nature gives rise to an FFLO state beyond the multi-band models.

In this work, we investigate superconducting SrTiO$_3$ in a parallel magnetic field on the basis of 
the three-orbital model, which reproduces the band structure of two-dimensional electron gases, 
and examine the roles of the orbital degree of freedom and spin-orbit coupling. 
It is shown that the orbital degeneracy in $t_{\rm 2g}$-orbitals on the Ti ions markedly affects the FFLO state. 
In particular, we obtain the following results for the multi-orbital FFLO state, 
(1) a rich phase diagram involving the orbital-dependent complex stripe state, 
which has not been revealed for single-band models,~\cite{SovPhysJETP.68.1244,JETPLett.78.637,PhysRevB.70.104521,PhysRevLett.94.137002,PhysRevB.75.064511,PhysRevB.76.014522} 
and (2) highly anisotropic behaviors of the superconducting state with respect to the in-plane rotation of the magnetic field. 
We also investigate the evolution of the FFLO state by increasing the carrier density, 
which can be controlled by applying the gate voltage.~\cite{Ueno,Ueno_review,Caviglia,Bell,Shalom,Caviglia-2}

In Sect.~2, we introduce the three-orbital tight-binding model for SrTiO$_3$ heterostructures, and explain 
the linearized gap equation by which the superconducting instability is investigated. 
In Sect.~3.1, the in-plane anisotropy of the critical magnetic field is shown and attributed to 
the spin texture, which is affected by the orbital degeneracy.  
The FFLO states in magnetic fields parallel to the [100]-axis and [110]-axis are studied in Sects.~3.2 and 3.3, respectively. 
The in-plane field angle dependence of the superconducting state is illustrated in Sect.~3.4. 
The carrier density dependence of the FFLO states is discussed in Sect.~4. 
Finally in Sect.~5, we summarize the results and propose experimental searches for the various FFLO states 
in SrTiO$_3$ heterostructures.

\section{Formulation}

\subsection{Three-orbital model for SrTiO$_3$ heterostructures}

We adopt a three-orbital tight-binding model for $t_{\rm 2g}$-orbitals on Ti ions. 
As shown by experiments~\cite{santander2011two,joshua2012universal,Berner,King2013} 
and electronic structure calculations,~\cite{King2013,delugas2011spontaneous,pentcheva2008ionic,
popovic2008origin,khalsa2012theory,hirayama2012ab,zhong2013theory}  
the conduction bands in SrTiO$_3$ heterostructures mainly consist of $t_{\rm 2g}$-orbitals. 
Although the electronic structure depends on the interface/surface termination, 
the band structure of two-dimensional electron gases is described by the one-body part of the Hamiltonian, 
\begin{align}
\hspace*{-3mm} 
H_{\rm 0} &= H_{\rm kin}+H_{\rm hyb}+H_{\rm CEF}+H_{\rm odd}+H_{\rm LS},\\
\hspace*{-3mm}
H_{\rm kin} &= \sum_{{\bm k}}\sum_{m=1,2,3}\sum_{s=\us,\ds} (\varepsilon_{m}({\bm k})-\mu) \,
c^{\dagger}_{{\bm k}, \, ms}c_{{\bm k}, \, ms},\\
\hspace*{-3mm}
H_{\rm hyb} &= \sum_{{\bm k}}\sum_{s=\us,\ds}[V({\bm k}) \, c^{\dagger}_{{\bm k}, \, 1s}c_{{\bm k}, \, 2s} + {\rm h.c.}],\\
\hspace*{-3mm}
H_{\rm CEF} &= \Delta \sum_{i}n_{3i},\\
\hspace*{-3mm}
H_{\rm odd} &= \sum_{{\bm k}}\sum_{s=\us,\ds}
[V_{\rm x}({\bm k}) \, c^{\dagger}_{{\bm k},\, 1s}c_{{\bm k},\, 3s}
+ V_{\rm y}({\bm k}) \, c^{\dagger}_{{\bm k}, \, 2s}c_{{\bm k}, \, 3s} + {\rm h.c.}],\\
\hspace*{-3mm} 
H_{\rm LS} &= \lambda\sum_{i}{\bm L}_{i} \cdot {\bm S}_{i}, 
\end{align}
where $c_{{\bm k},ms}$ is the annihilation operator for an electron with momentum ${\bm k}$, orbital $m$, and spin $s$. 
Here, the (d$_{yz}$, d$_{zx}$, d$_{xy}$)-orbitals are denoted by the orbital index $m=(1,2,3)$, respectively.
The first term $H_{\rm kin}$ describes the kinetic energy of each orbital and includes the chemical potential $\mu$.
The second term $H_{\rm hyb}$ is the intersite hybridization term of d$_{yz}$- and  d$_{zx}$-orbitals. 
The third term $H_{\rm CEF}$ introduces the crystal electric field with tetragonal symmetry.

Since the mirror symmetry with respect to the conducting plane is broken by the interface/surface, hybridization  
is allowed between the d$_{xy}$-orbital and (d$_{yz}$, d$_{zx}$)-orbitals,~\cite{zhong2013theory,Khalsa2013} 
which is represented by the odd-parity hybridization term $H_{\rm odd}$. 
Thus, the broken inversion symmetry is introduced in our model.  
The Rashba spin-orbit coupling appears as an effective spin-orbit coupling arising from the odd-parity hybridization 
term $H_{\rm odd}$ and the LS coupling term $H_{\rm LS}$~\cite{zhong2013theory,Yanase_NCSC,yanase2013electronic}. 
Although the Rashba spin-orbit coupling is often assumed phenomenologically in theoretical models~\cite{NCSC}, 
it is microscopically derived in our model. 

The following tight-binding forms are adopted in this paper.  
\begin{align}
\hspace{0mm}
\varepsilon_{\rm 1}({\bm k})&=-2t_{\rm 3}\cos k_{\rm x}-2t_{\rm 2}\cos k_{\rm y},\\
\hspace{0mm}
\varepsilon_{\rm 2}({\bm k})&=-2t_{\rm 2}\cos k_{\rm x}-2t_{\rm 3}\cos k_{\rm y},\\
\hspace{0mm}
\varepsilon_{\rm 3}({\bm k})&=-2t_{\rm 1}(\cos k_{\rm x}+\cos k_{\rm y})-4t_{\rm 4}\cos k_{\rm x}\cos k_{\rm y},\\
\hspace{0mm}
V({\bm k})&=4t_{\rm 5}\sin k_{\rm x}\sin k_{\rm y},\\
\hspace{0mm}
V_{\rm x}({\bm k})&=2{\rm i}t_{\rm odd}\sin k_{\rm x},\\
\hspace{0mm}
V_{\rm y}({\bm k})&=2{\rm i}t_{\rm odd}\sin k_{\rm y}.
\end{align}
The electronic structure of two-dimensional electron gases in the SrTiO$_3$/LaAlO$_3$ interface~\cite{
Berner,delugas2011spontaneous,pentcheva2008ionic,popovic2008origin,joshua2012universal,
khalsa2012theory,hirayama2012ab,zhong2013theory} is reproduced by the parameter set 
$
(t_{\rm 1}, t_{\rm 2}, t_{\rm 3}, t_{\rm 4}, t_{\rm 5}, t_{\rm odd}, \lambda, \Delta)
=(1, 1, 0.2, 0.4, 0.1, 0.1, 0.1, 2.45),
$
where the unit of energy is chosen as $t_{\rm 1}=1$. 
A first-principles band structure calculation~\cite{hirayama2012ab} obtained $t_{\rm 1} \simeq 300$ meV. 
In this paper, we mainly investigate a high two-dimensional carrier density of $n = 0.15 \simeq 1 \times 10^{14} \, {\rm cm}^{-2}$, 
which has been realized in a gate-controlled SrTiO$_3$/LaAlO$_3$ interface~\cite{Shalom} and in a SrTiO$_3$ surface.~\cite{Ueno} 
We show the carrier density dependence of the superconducting state in Sect.~4 and demonstrate that unusual properties disappear 
in the lower carrier density region.

For the study of superconductivity, we assume attractive interactions in the $s$-wave channel, 
\begin{align}
H_{\rm I}=&U\sum_{i}\sum_{m} n_{i,m\us} \, n_{i,m\ds} + U'\sum_{i}\sum_{m\neq m'} n_{i,m\us} \, n_{i,m'\ds}, 
\end{align}
where $U$ and $U'$ describe the intraorbital and interorbital attractive interactions, respectively. 
The $s$-wave symmetry of the superconductivity has been evidenced by both theory and experiment. 
A measurement of superfluid density showed the full excitation gap in the superconducting state.~\cite{bert2012gate} 
Furthermore, it has been theoretically shown that the transition temperature of approximately $0.3$ K and 
its nonmonotonic carrier density dependence in the SrTiO$_3$/LaAlO$_3$ interface are reproduced 
by the $s$-wave attractive interaction mediated by optical phonons.~\cite{Klimin} 
The renormalization of the band structure observed by ARPES measurements has also been explained by taking into account 
the electron-phonon coupling.~\cite{King2013} 
Thus, $s$-wave superconductivity is likely to occur in SrTiO$_3$ heterostructures as well as in bulk SrTiO$_3$.~\cite{Schooley} 
We assume $U=U'$ and choose $U$ so that the transition temperature at zero magnetic field is $T_{\rm c} = 0.001$. 
It has been confirmed that the following results are almost independent of $U$ and the ratio $U'/U$ 
as long as the reasonable condition $U'/U \leq 1$ is satisfied.~\cite{Nakamura_SCES} 
We do not take into account the pairing interaction in the odd-parity channel, and thus the parity mixing 
in Cooper pairs is ignored. 
This simplification is also justified because the role of induced odd-parity Cooper pairs in FFLO superconductivity 
is negligible unless the amplitude of odd-parity Cooper pairs is comparable to that of even-parity 
Cooper pairs.~\cite{JPSJ.76.124709,Comment1} 
Although superconductivity induced by odd-parity Cooper pairs has been theoretically proposed~\cite{Yada,Schmalian}, 
we do not consider such unconventional Cooper pairing, which may be caused by the strong electron correlation.

The purpose of our study is to elucidate the spin-polarized FFLO superconducting states 
in the two-dimensional electron gases formed on SrTiO$_3$ heterostructures. 
Thus, we consider the Zeeman coupling term, which is induced by a magnetic field or by the coexisting 
ferromagnetic order,~\cite{Michaeli} 
\begin{align}
H_{\rm Z}=&-\sum_{{\bm k}}\sum_{m}\sum_{s,s'}\mu_{\rm B}{\bm H}\cdot{\bm \sigma}_{ss'} \, 
c^{\dagger}_{{\bm k}, \, ms} c_{{\bm k}, \, ms'}, 
\end{align}
where ${\bm \sigma}$ is the Pauli matrix and $\mu_{\rm B}$ is the Bohr magneton. 
When we apply a magnetic field, superconductivity is destroyed through the orbital depairing effect 
as well as through the spin polarization induced by the Zeeman coupling term (paramagnetic depairing effect). 
However, the orbital depairing effect is negligible for a magnetic field parallel to two-dimensional electron gases,  
although it reduces the critical magnetic field along the [001]-axis to $H_{\rm c2}^{c} \sim 0.1$ T, 
which is much smaller than the Pauli limit. 
The ferromagnetic moment coexisting with the superconductivity is parallel to the conducting plane,~\cite{Dikin,Li,Bert,Ariando} 
and thus the proximity effect gives rise to the effective magnetic field ${\bm H}$ along the {\it ab}-plane. 
Thus, we assume a parallel magnetic field in the following sections and ignore the orbital depairing effect. 
The total Hamiltonian is given by 
%
$
H=H_{\rm 0} + H_{\rm I} + H_{\rm Z}. 
$

\subsection{Linearized gap equation}

We determine the instability to the superconducting state by solving the linearized mean field 
gap equation formulated in the following manner.
First, we diagonalize the one-body Hamiltonian $H_{\rm 0} + H_{\rm Z}$ using the unitary matrix 
\begin{eqnarray}
\hat{U}({\bm k}) = \left(
\begin{array}{cccc}
u_{1\us,1} & \cdots & u_{1\us,6}     \\
u_{2\us,1} & \cdots & u_{2\us,6}     \\
\vdots & \ddots & \vdots                    \\
u_{3\ds,1} & \cdots & u_{3\ds,6}  \\
\end{array}
\right). 
\end{eqnarray}
Thereby, the basis changes as $C^{\dagger}_{{\bm k}}=\Gamma^{\dagger}_{{\bm k}}U^{\dagger}({\bm k})$, 
where 
$C^{\dagger}_{{\bm k}}=(c^{\dagger}_{{\bm k}, \, 1\us},c^{\dagger}_{{\bm k}, \, 2\us},\cdots ,
c^{\dagger}_{{\bm k}, \, 3\ds})$
and 
$\Gamma^{\dagger}_{{\bm k}}=(\gamma^{\dagger}_{{\bm k}, \, 1},\gamma^{\dagger}_{{\bm k}, \, 2},\cdots ,
\gamma^{\dagger}_{{\bm k}, \, 6})$.  
By using the operators in the band basis, 
the one-body Hamiltonian is diagonalized as 
\begin{eqnarray}
H_{\rm 0}+H_{\rm z}=\sum_{{\bm k}} \sum_{j=1}^{6} E_{j}({\bm k}) \ \gamma^{\dagger}_{{\bm k}, \, j} \ \gamma_{{\bm k}, \, j},
\end{eqnarray}
where $E_{j}({\bm k})$ is the energy of a quasiparticle and we take $E_{i}({\bm k}) \geq E_{j}({\bm k})$ for $i > j$.
The Matsubara Green functions in the spin and orbital basis are obtained as@
\begin{align}
G_{m's', \ ms}({\bm k},{\rm i}\omega_{l})&=-\int_{0}^{\beta}d\tau e^{{\rm i}\omega_{l}\tau}\langle 
c_{{\bm k}, \, m's'}(\tau)c^{\dagger}_{{\bm k}, \, ms}(0) \rangle, \\
=&\sum^{6}_{j=1}\frac{1}{{\rm i}\omega_{l}-E_{j}({\bm k})}u_{m's', j}({\bm k})u^{*}_{ms, j}({\bm k}), 
\end{align}
where $\omega_{l}$ is the Matsubara frequency. 

The linearized gap equation is formulated by considering the divergence of the T-matrix, which is defined as 
\begin{eqnarray}
&& \hspace*{-15mm}
\hat{T}({\bm q}) = \hat{T}_{\rm 0}({\bm q}) - \hat{T}({\bm q}) \, \hat{H}_{\rm I} \, \hat{T}_{\rm 0}({\bm q}).   
\end{eqnarray}
The wave vector ${\bm q}$ represents the center-of-mass momentum of Cooper pairs. 
The matrix element of the irreducible T-matrix $\hat{T}_{0}({\bm q})$ is obtained as 
\begin{eqnarray}
&& \hspace{-14mm}
T_{\rm 0}^{(m n, \, m' n')}({\bm q}) 
\nonumber \\ &&  \hspace{-14mm}
=T \sum_{\omega_{l}}\sum_{{\bm k}}
[ G_{m'\us, \, m\us}({\bm q}/2+{\bm k},{\rm i}\omega_{l}) 
G_{n'\ds, \, n\ds}({\bm q}/2-{\bm k},-{\rm i}\omega_{l})    
\nonumber \\
&& 
-G_{n'\ds, \, m\us}({\bm q}/2+{\bm k},{\rm i}\omega_{l}) 
G_{m'\us, \, n\ds}({\bm q}/2-{\bm k},-{\rm i}\omega_{l}) ], 
\end{eqnarray}
where $T$ is the temperature. 
When we represent the T-matrix using the basis $(mn)=(11,12,13,21,22,23,31,32,33)$, 
the interaction term is represented by the $9 \times 9$ diagonal matrix  
$\hat{H}_{\rm I} = \left( U_{m} \delta_{mn} \right)$ with $U_{m} = U$ for $m=1, 5, 9$ and 
$U_{m}=U'$ for others. 
The superconducting transition occurs when the maximum eigenvalue $\lambda_{\rm max}$ of the matrix, 
$ - \hat{H}_{\rm I} \hat{T}_{\rm 0}(\q)$, is unity. 
The order parameter in the orbital basis is obtained from the eigenvector $(\psi_{mn})$, which is proportional to 
$\Delta_{mn} = - g \sum_{k} \langle c_{{\bm q}/2 + {\bm k}, \, m \us} c_{{\bm q}/2 - {\bm k}, \, n \ds} \rangle$ with 
$g=U$ for $m=n$ and $g=U'$ for $m \ne n$. 
Thus, the superconducting state below the transition temperature $T_{\rm c}(H)$ is determined by the linearized gap equation, 
although the full solution of the Bogoliubov-de Gennes equation is required for studying the superconducting state at 
low temperatures.

The superconducting transition is induced by the quasi-long-range order in two-dimensional systems 
and described by the Berezinskii-Kosterlitz-Thouless (BKT) transition.~\cite{Berezinskii,KT} 
Although even the quasi-long-range order of FFLO superconductivity is suppressed 
in isotropic two-dimensional systems, the anisotropy in the Fermi surface allows the quasi-long-range order.~\cite{Shimahara_BKT} 
Indeed, critical behaviors of the BKT transition have been observed 
in the SrTiO$_3$/LaAlO$_3$ interface.~\cite{Reyren,Caviglia,Schneider} 
The BKT transition temperature is roughly obtained using the following equation:~\cite{KT} 
\begin{eqnarray}
T_{\rm BKT} = \frac{\pi}{2 m^*} \, \rho\left(T_{\rm BKT}\right), 
\end{eqnarray}
where $\rho\left(T\right)$ is the superfluid density. 
Because of the large superfluid density and small transition temperature of SrTiO$_3$ heterostructures,  
the transition temperature obtained on the basis of mean field theory ($T_{\rm c}$) almost coincides with the 
BKT transition temperature ($T_{\rm BKT}$),~\cite{Reyren} as $1-T_{\rm BKT}/T_{\rm c} \ll 1$.  
Thus, the superconductivity in SrTiO$_3$ heterostructures is approximately described on the basis of mean field theory, 
although the long-range order obtained using mean field theory is interpreted as a quasi-long-range order at finite temperatures.

\subsection{Classification of FFLO states} 

For clarity, we classify various FFLO states, which will appear in the following sections.  
Although the FFLO states have been classified into the single-$Q$ FF state~\cite{FF} and double-$Q$ LO state,~\cite{LO}
the so-called helical state can emerge in noncentrosymmetric superconductors.~\cite{
SovPhysJETP.68.1244,JETPLett.78.637,PhysRevB.70.104521,PhysRevLett.94.137002,PhysRevB.75.064511,PhysRevB.76.014522} 
The order parameter of the helical state is formally the same as that of the FF state, 
but the magnitude of the center-of-mass momentum, 
$|\q| \sim  \left(\alpha/E_{\rm F}\right) \cdot \left(\mu_{\rm B}H/T_{\rm c}\right) \cdot \xi^{\,-1}$, is much smaller than 
that in the FF state, $|\q| \sim \xi^{\,-1}$, where $\xi$ is the coherence length. 
The helical state is stabilized by the asymmetric deformation of Fermi surfaces arising from 
the antisymmetric spin-orbit coupling and magnetic field.~\cite{NCSC} 
Therefore, the magnitude $|\q|$ is proportional to $\alpha$ and $H$. 
Thus, long-period spatial modulation characterizes the helical state. 
Indeed, the crossover from the helical state to the FF state occurs by increasing the magnetic field,~\cite{JPSJ.76.124709} 
and we differentiate these two states by looking at the magnitude of $\q$. 
Above the crossover field, the CS state where 
$\Delta(\rr) = \Delta \left(e^{i \q \rr} + \delta e^{-i \q \rr}\right)/2 \ne \Delta \cos(\q \rr) $ 
may be stabilized in the low-temperature region.~\cite{PhysRevB.75.064511} 
Below, we will show that a novel two-component CS state due to the orbital degree of freedom is stabilized 
in SrTiO$_3$ heterostructures (see Sect.~3.3).

\section{FFLO superconductivity}

As we find that the superconducting state markedly depends on the direction of the magnetic field,  
we first show the anisotropic paramagnetic depairing effect in Sect.~3.1. Then, we show 
the FFLO superconducting states for $\HH \parallel $ [100] and $\HH \parallel $ [110] 
in Sects.~3.2 and 3.3, respectively. The field angle dependence of the superconducting state 
for $\HH \parallel (\cos \theta, \sin \theta, 0)$ is discussed in Sect.~3.4.

\subsection{In-plane anisotropy of paramagnetic depairing effect}

\begin{figure}[ht]
\centering
\includegraphics[width=7.0cm,origin=c,keepaspectratio,clip]{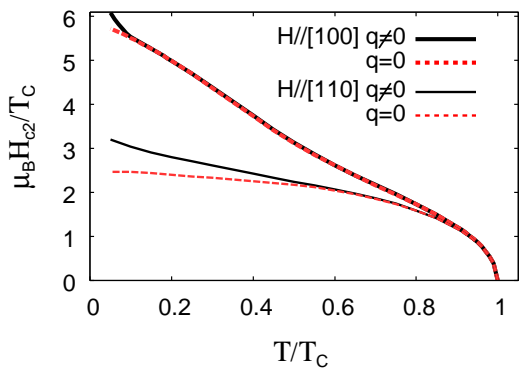}
\caption{(Color online) 
Critical magnetic fields of the BCS state (red dashed lines) and FFLO state (black solid lines). 
Thick and thin lines show the normalized critical magnetic fields $\mu_{\rm B}H_{\rm c2}/T_{\rm c}$ along the [100]- 
and [110]-axes, respectively. 
We assume the carrier density $n = 0.15 \simeq 1 \times 10^{14} \, {\rm cm}^{-2}$ throughout Sect.~3. 
}
\label{Fig.2}
\end{figure}

First, we show the anisotropy of the paramagnetic depairing effect, which arises from the orbital degeneracy 
in the electronic structure. 
We also study the thermodynamical stability of the FFLO state by comparing the critical magnetic fields 
of the BCS state and FFLO state. 
Figure~1 shows that the critical magnetic field is enhanced by allowing the Cooper pairs to have a finite center-of-mass momentum. 
However, for $\HH \parallel$ [100], the enhancement is negligible and is much smaller than the effect of 
canonical FFLO superconductivity on the critical magnetic field.~\cite{FFLO_review} 
On the other hand, the critical magnetic field along the [110]-axis is considerably enhanced. 
Since the critical magnetic field for $\HH \parallel$ [110] is smaller than that for $\HH \parallel$ [100], 
a larger paramagnetic depairing effect is indicated for the former.

\begin{figure}[ht]
\centering
\includegraphics[width=5.0cm,origin=c,keepaspectratio,clip]{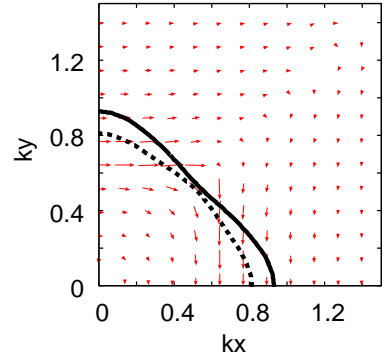}
\caption{(Color online) 
Spin texture in the 1st band. 
Arrows show the g-vector $\g_1({\bm k})$ defined in Eq.~(\ref{g-vector}). 
Solid and dashed lines show the Fermi surfaces split by spin-orbit coupling. 
}
\label{Fig.3}
\end{figure}

We explain the large anisotropy of the paramagnetic depairing effect by illustrating the spin texture 
of spin-split bands. 
The spin degeneracy in the band structure is lifted by the spin-orbit coupling in a noncentrosymmetric metal.~\cite{NCSC} 
The momentum-dependent spin polarization is described by the ``g-vector'', which is defined for the $l$th band 
as~\cite{yanase2013electronic,Nakamura_STO} 
\begin{eqnarray}
\label{g-vector}
&& \hspace{-10mm}
\g_l({\bm k}) = \left(E_{2l}({\bm k})-E_{2l-1}({\bm k})\right) \tilde{\SS}_{2l}^{\rm \ av}({\bm k}). 
\end{eqnarray} 
The spin polarization axis 
$\tilde{\SS}_{2l}^{\rm \ av}({\bm k}) = \SS_{2l}^{\rm \ av}({\bm k})/|\SS_{2l}^{\rm \ av}({\bm k})|$ 
is obtained by calculating the average spin 
$\SS_{2l}^{\rm \ av}({\bm k}) = \langle \sum_{m}\sum_{ss'} 
{\bm \sigma}_{ss'} c_{{\bm k}, \, m s}^{\dag}c_{{\bm k}, \, m s'} \rangle_{2l}$ for each momentum $\k$. 
The arrows in Fig.~2 show the g-vector of the 1st band whose spin-split Fermi surfaces are shown by the 
solid and dashed lines. We see that the g-vector is almost perpendicular to the [100]-axis 
in half of the Brillouin zone, $|k_x| > |k_y|$, where the 1st band mainly consists of the d$_{yz}$-orbital. 
As shown in the literature,~\cite{NCSC} Cooper pairs are robust against the paramagnetic depairing effect when 
the spin polarization axis is perpendicular to the magnetic field. Thus, the quasi-one-dimensional 
superconducting state mainly induced by the d$_{yz}$-orbital substantially avoids the paramagnetic depairing effect 
for $\HH \parallel$ [100].~\cite{Nakamura_STO}. 
Indeed, a large critical magnetic field beyond the Pauli limit has been observed in the SrTiO$_3$/LaAlO$_3$ interface.~\cite{Shalom} 
On the other hand, we see that the spin polarization axis is not perpendicular to the [110]-axis in the entire 
Brillouin zone except for a tiny region near $k_x = k_y$.
Therefore, the large paramagnetic depairing effect suppresses the BCS state for $\HH \parallel$ [110], and it is partly avoided 
in the FFLO state.

The large anisotropy discussed above is attributed to the orbital degree of freedom in the $t_{2g}$ electron system. 
Indeed, the spin texture shown in Fig.~2 is typical for an orbitally degenerate noncentrosymmetric metal. 
When the crystal electric field is sufficiently large so that the orbital degree of freedom is quenched, 
an often assumed Rashba-type antisymmetric spin-orbit coupling [$\g_l({\bm k}) \propto (\sin k_y, - \sin k_x, 0)$] 
is obtained.~\cite{yanase2013electronic} 
On the other hand, in orbitally degenerate systems, the g-vector dramatically changes direction 
near the symmetric axis in the Brillouin zone,~\cite{yanase2013electronic} as clearly shown in Fig.~2. 
Thus, the characteristic spin texture in the multi-orbital systems causes the anisotropic behaviors of superconductivity 
with respect to the magnetic field. 
In the following subsections, we show that the FFLO states in SrTiO$_3$ heterostructures depend on 
the in-plane direction of the magnetic field. 

\subsection{Nonmonotonic evolution of FFLO state in $\HH \parallel$ [100]}

\begin{figure}[ht]
\centering
\includegraphics[width=7.0cm,origin=c,keepaspectratio,clip]{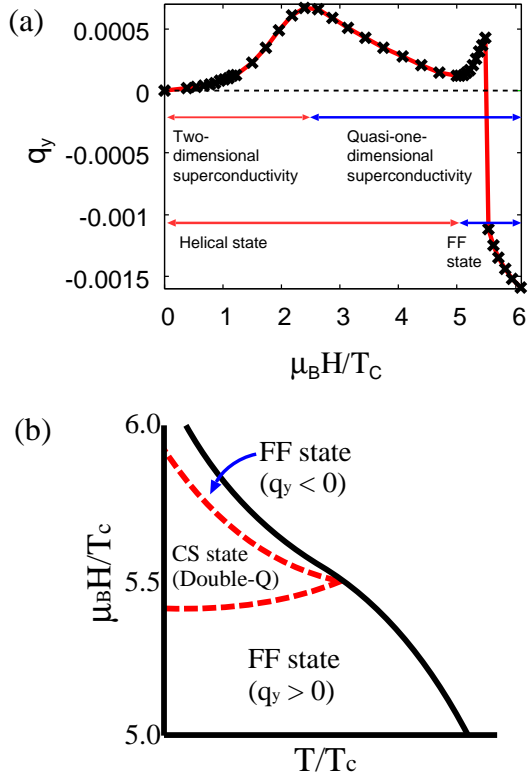}
\caption{(Color online) 
(a) center-of-mass momentum in Cooper pairs at $T=T_{\rm c}(H)$ as a function of the magnetic field along the [100]-axis. 
As $\q = q_y \hat{y}$, we plot $q_y$. The crossover in the superconducting state is described (see the text for details). 
(b) Schematic phase diagram in the high-magnetic-field region. Multiple superconducting transitions 
are illustrated by red dashed lines. 
}
\label{Fig.1}
\end{figure}

The superconducting state shows several crossovers and the FFLO state nonmonotonically changes 
with increasing magnetic field along the [100]-axis. 
The antisymmetric spin-orbit coupling arising from the interfacial mirror-symmetry breaking is of the Rashba type,~\cite{Rashba} 
and thus the Cooper pairs acquire a center-of-mass momentum perpendicular to the magnetic field.~\cite{NCSC} 
Indeed, we obtain $\q \parallel$ [010] unless $\HH =0$. 
Figure~3(a) shows the [010]-component $q_{y}$ as a function of the magnetic field. 
We see a peak at approximately $\mu_{\rm B}H/T_{\rm c} = 2.4$, 
although $|q_{y}|$ monotonically increases with the magnetic field in single-band models.~\cite{JPSJ.76.124709} 
The peak is associated with the dimensional crossover of the superconducting state.~\cite{Nakamura_STO} 
The superconductivity is mainly induced by the degenerate (d$_{yz}$, d$_{zx}$)-orbitals in the low-magnetic-field region, 
and it changes to the quasi-one-dimensional superconducting state induced by the d$_{yz}$-orbital 
at approximately $\mu_{\rm B}H/T_{\rm c} = 2.4$. 
Since the paramagnetic depairing effect is suppressed in the latter, as we discussed in Sect.~3.1, 
the center-of-mass momentum in Cooper pairs is decreased with increasing magnetic field 
for $2.4 < \mu_{\rm B}H/T_{\rm c} < 5$.

The superconducting state shows another crossover from the helical state to the FF state at approximately $\mu_{\rm B}H/T_{\rm c} = 5$, 
as indicated by the second increase in $q_{y}$ above $\mu_{\rm B}H/T_{\rm c} > 5$. 
Then, the sign of $q_{y}$ suddenly changes at $\mu_{\rm B}H/T_{\rm c} = 5.5$. 
Because the two momenta are degenerate at $\mu_{\rm B}H/T_{\rm c} = 5.5$, the multiple superconducting phases appear, 
as illustrated in Fig.~3(b).
Here, it is assumed that the Cooper pair condensates for both momenta $q_{y} = q_+ \sim + 0.0005$ and $q_{y} = q_- \sim - 0.0011$ 
coexist below $T_{\rm c}$ as in the single-band model.~\cite{PhysRevB.75.064511} 
The superconducting state shows a phase transition from the single-$Q$ FF state with $q_{y} > 0$ to the double-$Q$ state, 
and it again changes to the single-$Q$ FF state with $q_{y} < 0$ with increasing magnetic field. 

The order parameter in the double-$Q$ state is described as 
\begin{eqnarray}
\Delta_{mn}(\rr) = \Delta_{mn}^{(+)} e^{i q_+ y} + \Delta_{mn}^{(-)} e^{i q_- y}.
\end{eqnarray}
Since $|q_{+}| \ne |q_{-}|$, both the amplitude and phase of the order parameter are spatially inhomogeneous as in the CS state. 
As the sign change of $q_{y}$ does not occur in single-orbital models,~\cite{JPSJ.76.124709,PhysRevLett.94.137002,PhysRevB.75.064511}
the multiple superconducting transitions illustrated in Fig.~3(b) are attributed to the orbital degree of freedom in $t_{2g}$ electron systems.

\subsection{Orbital-dependent FFLO state in $\HH \parallel$ [110]}

Next, we investigate the FFLO state in a magnetic field along the [110]-axis. 
Figure~4 shows the marked increase in the center-of-mass momentum in Cooper pairs above the magnetic field 
$\mu_{\rm B}H/T_{\rm c} >1.7$. Similar behavior has been obtained for single-band models for noncentrosymmetric 
superconductors.~\cite{JPSJ.76.124709} It indicates the crossover from the helical state to the FF state. 
However, the superconducting state in the high-magnetic-field region is distinct from the conventional FF state, 
as we will show below.

\begin{figure}[ht]
\centering
\includegraphics[width=6.0cm,origin=c,keepaspectratio,clip]{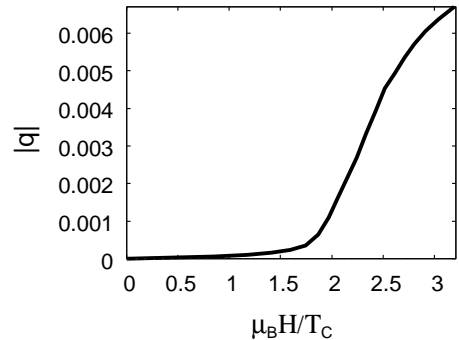}
\caption{
Magnitude of center-of-mass momentum $|\q|$ at $T=T_{\rm c}(H)$ as a function of the magnetic field along the [110]-axis. 
}
\label{Fig.4}
\end{figure}

Figure~5 shows the center-of-mass momentum $\q=(q_x, q_y)$ in the two-dimensional momentum space for various magnetic fields. 
Although the isotropic Rashba superconductor acquires a center-of-mass momentum perpendicular to the in-plane magnetic field, 
namely, $\q \parallel [1\bar{1}0]$ in this case, we see a deviation of $\q$ from the symmetric $[1\bar{1}0]$-axis 
at high magnetic fields, $\mu_{\rm B}H/T_{\rm c} > 2.3$. 
The deviation is caused by the quasi-one-dimensional nature of the Fermi surface, which favors the FFLO state 
with $\q \parallel [100]$ or $\q \parallel [010]$. 
The high-field FFLO state is determined by the competition between the Rashba spin-orbit coupling and 
the anisotropy in Fermi surfaces, and the center-of-mass momentum is generally represented as  
$\q_{1,2} = \q_\perp \pm \q_\parallel$. 
Here, we decomposed $\q_{1}$ and $\q_{2}$ into $\q_\perp \parallel [1\bar{1}0]$ and $\q_\parallel \parallel [110]$.

\begin{figure}[ht]
\centering
\includegraphics[width=8.0cm,origin=c,keepaspectratio,clip]{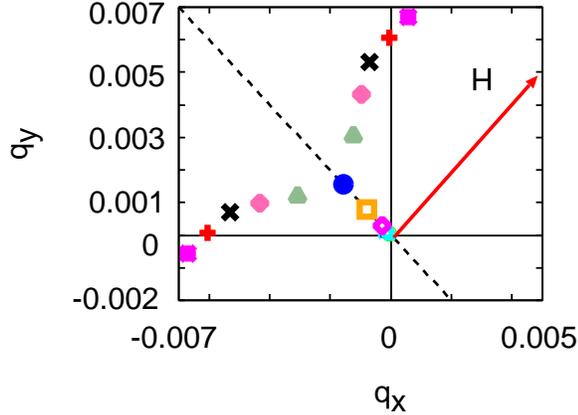}
\caption{(Color online) 
Center-of-mass momentum of Cooper pairs for various magnetic fields along the [110]-axis. 
Each mark is obtained at a magnetic field and temperature represented in Fig.~6. 
The superconducting state is described by the two-component order parameters with respect to the center-of-mass momentum 
in the high-magnetic-field region. 
}
\label{Fig.5}
\end{figure}

Because of the mirror symmetry along the [110]-axis, the superconducting state has twofold degeneracy  
with respect to the center-of-mass momentum, $\q_{1}$ and $\q_{2}$. 
Thus, the superconducting state is described by the two-component 
order parameters $\bm{\eta}=(\eta_{1}, \eta_{2})$, where $\eta_{1}$ corresponds to $\q_{1}$ and $\eta_{2}$ corresponds to $\q_{2}$. 
The Ginzburg-Landau free energy is described as  
\begin{eqnarray}
&& \hspace{-10mm}
F = -\alpha_0 \left(1-\frac{T}{T_{\rm c}(H)}\right) \, | \bm{\eta} |^2 
+ \frac{\beta}{2} \, | \bm{\eta} |^4 + \gamma \, | \eta_{1} |^2 \, | \eta_{2} |^2
\nonumber \\ && \hspace{-7mm}
+ \kappa_1 \left(|D_{\tilde{x}} \, \eta_{1}|^2 + |D_{\tilde{x}'} \, \eta_{2}|^2 \right)
+ \kappa_2 \left(|D_{\tilde{y}} \, \eta_{1}|^2 + |D_{\tilde{y}'} \, \eta_{2}|^2 \right),  
\label{GL}
\end{eqnarray}
where $D_{a} = -i \partial_a + 2 e A_a$ is a covariant derivative. Note that the 
principal axis of gradient terms is different between the two-component order parameters, 
as $(\tilde{x}, \tilde{y}) \ne (\tilde{x}', \tilde{y}')$. 
When the magnetic field is precisely applied to the {\it ab}-plane, the gradient terms do not play any role. 
Then, the double-$Q$ state, where $\bm{\eta} \propto (1,e^{i \theta})$, is stable when $\gamma < 0$. On the other hand, 
the single-$Q$ state, where $\bm{\eta} \propto (0,1)$ or $\bm{\eta} \propto (1,0)$, is stable otherwise.

The order parameter in the orbital basis is represented by ${\bm \eta}$ as
\begin{eqnarray}
&& \hspace{-10mm}
\Delta_{mn}(\rr) = \eta_{1} \, \Delta_{mn}^{\rm (1)} \, e^{i \q_{1} \rr} + \eta_{2} \, \Delta_{mn}^{\rm (2)} \, e^{i \q_{2} \rr}, 
\\ && \hspace{0mm} 
= e^{i \q_{\perp} \rr} \left[ \eta_{1} \, \Delta_{mn}^{\rm (1)} \, e^{i \q_{\parallel} \rr} + \eta_{2} \, \Delta_{mn}^{\rm (2)} \, e^{-i \q_{\parallel} \rr} \right],
\end{eqnarray}
where $\Delta_{mn}^{\rm (1)}$ and $\Delta_{mn}^{\rm (2)}$ are obtained by the linearized gap equation for $\q = \q_{1}$ and $\q = \q_{2}$, 
respectively. 
For both momenta, the dominant components are $\Delta_{11}^{\rm (1,2)}$ and $\Delta_{22}^{\rm (1,2)}$, which describe 
the intra-orbital Cooper pairs formed by the d$_{\rm yz}$- and d$_{\rm zx}$-orbitals, respectively. 
We find that $|\Delta_{11}^{\rm (1)}| = |\Delta_{22}^{\rm (2)}| <  |\Delta_{11}^{\rm (2)}| = |\Delta_{22}^{\rm (1)}|$, 
because the d$_{\rm yz}$-orbital (d$_{\rm zx}$-orbital) favors the FFLO state with $\q \parallel [100]$ ($\q \parallel [010]$). 
Thus, the Cooper pairs in the single-$Q$ state are ``orbital-polarized''.

On the other hand, the order parameter in the double-$Q$ state is described as 
\begin{eqnarray}
&& \hspace{-10mm}
\Delta_{11}(\rr) = \Delta \, e^{i \q_{\perp} \rr} \left[\delta \, e^{i \q_{\parallel} \rr} + e^{-i \q_{\parallel} \rr} \right], 
\label{double-q-11}
\\ && \hspace{-10mm}
\Delta_{22}(\rr) = \Delta \, e^{i \q_{\perp} \rr} \left[e^{i \q_{\parallel} \rr} + \delta \, e^{-i \q_{\parallel} \rr} \right], 
\label{double-q-22}
\end{eqnarray}
where $\Delta = |\Delta_{11}^{\rm (2)}| = |\Delta_{22}^{\rm (1)}|$, and $\delta < 1$. 
For instance, we obtain $\delta = 0.88$ at $\mu_{\rm B}H/T_{\rm c} = 3.2$. 
Equations (\ref{double-q-11}) and (\ref{double-q-22}) are similar to the order parameter in the CS state, 
but they acquire a phase oscillation $e^{i \q_{\perp} \rr}$. 
Furthermore, the average center-of-mass momentum in Cooper pairs depends on the orbital: 
it is $\q_{\perp} - \frac{1-\delta^2}{1+\delta^2} \, \q_{\parallel}$ for the d$_{yz}$-orbital and 
$\q_{\perp} + \frac{1-\delta^2}{1+\delta^2} \, \q_{\parallel}$ for the d$_{zx}$-orbital. 
Thus, we call the double-$Q$ state the ``orbital-dependent complex stripe (ODCS) state''. 
An analogous ``layer-dependent complex stripe state'' has been proposed for multilayer superconductors 
affected by spin-orbit coupling.~\cite{Yoshida_CS}

\begin{figure}[ht]
\centering
\includegraphics[width=7.0cm,origin=c,keepaspectratio,clip]{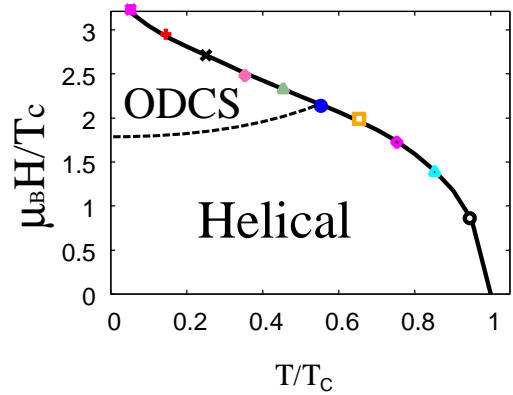}
\caption{(Color online) 
Phase diagram for $\HH \parallel$ [110]. 
The solid line is the critical magnetic field obtained by our calculation, and 
the marks correspond to those in Fig.~5. A second-order phase transition line is schematically drawn by the dashed line. 
The ODCS state and helical state are stabilized in the high- and low-magnetic-field regions, respectively. 
}
\label{Fig.6}
\end{figure}

Because the orbital polarization in Cooper pairs costs finite energy, 
it is expected that the double-$Q$ state is more stable than the single-$Q$ state. 
Since the superconducting gap in the single-$Q$ state is anisotropic in the momentum space, 
the single-$Q$ state gains less condensation energy than the double-$Q$ state. Thus, we assume $\gamma < 0$ and 
draw the schematic phase diagram in Fig.~6. 
The helical state is stabilized in the low-magnetic-field region, while the ODCS state is stable in the high-magnetic-field region. 
The continuous second-order phase transition occurs in the superconducting state, as illustrated by the dashed line, where 
$\q_{\parallel}$ grows with increasing magnetic field.

The Ginzburg-Landau free energy in Eq.~(\ref{GL}) conserves the $U(1) \times U(1)$ symmetry in the order parameter manifold. 
Therefore, the fractional vortex can emerge when the magnetic field is slightly tilted from the {\it ab}-plane. 
When the vortex cores of the $\eta_1$ component are shifted from those of the $\eta_2$ component, 
as favored for $- \beta < \gamma < 0$, a fractional vortex lattice is formed. 
The fractional vortex lattice has been topologically identified to be 
a skyrmion lattice of superconductivity.~\cite{Agterberg_skyrmion}

\subsection{Field angle dependence}

\begin{figure}[ht]
\centering
\includegraphics[width=8.0cm,origin=c,keepaspectratio,clip]{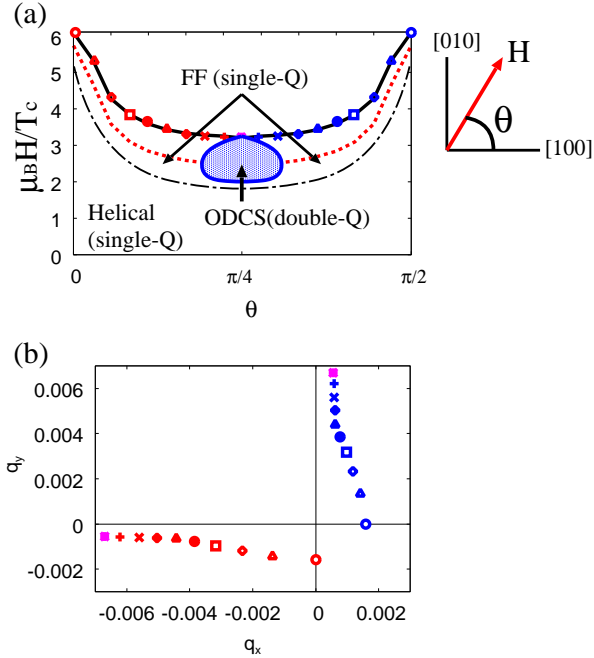}
\caption{(Color online) 
(a) The solid line shows the critical magnetic field at $T/T_{\rm c} = 0.05$ as a function of $\theta$, 
the angle of the in-plane magnetic field 
$\HH = H (\cos \theta, \sin \theta, 0)$. The red dashed line is the critical field of the BCS state. 
We also show the crossover line between the helical state and the FF state (black dash-dotted line). 
The ODCS phase is schematically drawn by the shaded area. 
The double-$Q$ CS state in Fig.~3(b) is not shown in this figure.  
(b) Cooper pair momentum $\q=(q_x, q_y)$ at $T/T_{\rm c} = 0.05$ and $H = H_{\rm c2}(T)$
for various magnetic field directions from $\theta =0$ to $\theta =\pi/2$. 
Marks in Fig.~7(b) correspond to those in Fig.~7(a). For example, the open blue circle is obtained for $\theta = \pi/2$. 
}
\label{Fig.7}
\end{figure}

As we have discussed in Sects.~3.1-3.3, the paramagnetic depairing effect and the resulting FFLO state are markedly different 
between $\HH \parallel $ [100] and $\HH \parallel $ [110] because of the orbital degree of freedom in the $t_{\rm 2g}$ electrons. 
Here, we clarify the field angle dependence of the superconducting state for $\HH \parallel (\cos \theta, \sin \theta, 0)$. 

Figure~7(a) shows the field angle dependence of the critical magnetic field. 
As we showed in Sect.~3.1, the critical magnetic field is enhanced near $\theta = 0$ and $\theta = \pi/2$, 
while it is insensitive to $\theta$ at around $\theta = \pi/4$. 
Thus, an experimental observation of the enhancement of the critical magnetic field at around $\theta = 0$ 
may indicate the orbital degree of freedom which plays an important role on the superconductivity 
in SrTiO$_3$ heterostructures.

When the magnetic field is slightly tilted from the [110]-axis, the degeneracy with respect to the center-of-mass momentum 
$\q = \q_{1}$ and $\q = \q_{2}$ is lifted. Therefore, the single-$Q$ FF state is stable just below the critical magnetic field. 
However, the ODCS state is stabilized by decreasing the magnetic field when the field angle is close to $\theta = \pi/4$. 
A schematic phase diagram is drawn in Fig.~7(a). 
The FF state continuously changes to the helical state with decreasing magnetic field, as we discussed earlier.

Figure~7(b) shows the field angle dependence of the Cooper pair momentum. 
For $\HH \parallel$ [100] and [010], the momentum $\q$ is perpendicular to the magnetic field, and the magnitude $|\q|$ is small 
because the paramagnetic depairing effect is suppressed. 
The magnitude grows with the tilting of the magnetic field to the [110]-axis. 
The center-of-mass momentum of the Cooper pairs markedly changes through $\theta = \pi/4$, where 
two momenta $\q = \q_{1}$ and $\q = \q_{2}$ are degenerate.

\section{Carrier Density Dependence of FFLO State}

\begin{figure}[ht]
\centering
\includegraphics[width=6.5cm,origin=c,keepaspectratio,clip]{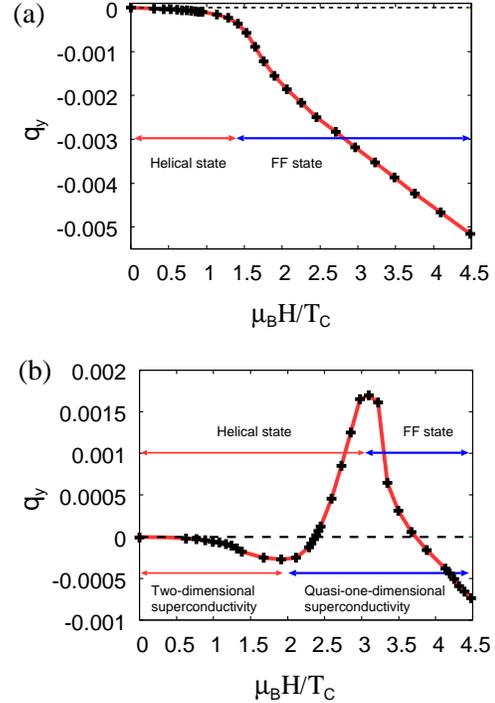}
\caption{(Color online) 
center-of-mass momentum in Cooper pairs at $T=T_{\rm c}(H)$ as a function of the magnetic field along the [100]-axis. 
We assume carrier densities (a) $n = 0.05 \simeq 3.5 \times 10^{13} \, {\rm cm}^{-2}$ and 
(b) $n = 0.1 \simeq 7 \times 10^{13} \, {\rm cm}^{-2}$. 
We plot $q_y$ as in Fig.~3(a). 
}
\label{Fig.8}
\end{figure}

While we clarified the unconventional FFLO states in the high-carrier-density region with 
$n = 0.15 \simeq 1 \times 10^{14}$ cm$^{-2}$, 
here, we show that they disappear in the low-carrier-density region. 
It has been shown that the superconducting state changes with increasing carrier density 
from the d$_{xy}$-orbital-induced superconductivity to that induced by d$_{yz}$/d$_{zx}$-orbitals.~\cite{Nakamura_STO}
For our choice of parameters, the crossover occurs at around $n=0.07 \simeq 5 \times 10^{13}$ cm$^{-2}$. 
This crossover density is larger than the experimentally observed value of 
$n \simeq 2 \times 10^{13}$ cm$^{-2}$,~\cite{joshua2012universal} but they are in reasonable agreement with each other.
Below the crossover density, the orbital degree of freedom is almost quenched, and thus 
the unusual properties of FFLO superconductivity discussed in Sect.~3 do not appear. 
Indeed, the Cooper pair momentum shows conventional growth with the magnetic field [Fig.~8(a)]. 
For $n=0.05 \simeq 3.5 \times 10^{13}$ cm$^{-2}$, the superconducting state shows a single 
crossover from the helical state to the FF state as in single-band models.~\cite{JPSJ.76.124709} 
Since the paramagnetic depairing effect is not suppressed as much as in the quasi-one-dimensional superconducting state for 
$n = 0.15 \simeq 1 \times 10^{14}$ cm$^{-2}$, the magnitude of the Cooper pair momentum is much larger than 
that in the high-carrier-density region [compare Fig.~8(a) with Fig.~3(a)]. 
We confirmed that the Cooper pair momentum is perpendicular to the magnetic field even for $\HH \parallel$ [110]. 
Thus, the ODCS state is not stabilized at $n=0.05 \simeq 3.5 \times 10^{13}$ cm$^{-2}$.

Near the crossover carrier density, namely, $n=0.1 \simeq 7 \times 10^{13}$ cm$^{-2}$, the center-of-mass momentum in Cooper pairs 
changes sign twice, corresponding to the two crossovers discussed in Sect.~3.2. 
$q_{\rm y}$ is negative in the two-dimensional superconducting state at low magnetic fields, but it is positive in 
the quasi-one-dimensional superconducting state in the intermediate magnetic field region. It again changes sign 
at $\mu_{\rm B}H/T_{\rm c} \simeq 3.7$, where the helical state changes to the FF state. 
Thus, the nonmonotonic magnetic field dependence appears in the center-of-mass momentum of Cooper pairs. 
However, the sudden change in $q_{\rm y}$ does not occur in contrast to that shown in Fig.~3(a). 
Therefore, the double-$Q$ CS state is not stabilized at least near $T=T_{\rm c}(H)$. 
When we apply the magnetic field along the [110]-axis, the ODCS state is stabilized, but the critical temperature of 
the ODCS state ($T_{\rm ODCS}/T_{\rm c} \sim 0.1$) is much smaller than that in the high-carrier-density region 
$n = 0.15 \simeq 1 \times 10^{14}$ cm$^{-2}$ ($T_{\rm ODCS}/T_{\rm c} \sim 0.5$). 
These results show that unusual FFLO states appear in SrTiO$_3$ heterostructures in the high-carrier-density region 
above the crossover density. 

\section{Summary and Discussion}

We have investigated the multi-orbital FFLO superconductivity in two-dimensional electron gases 
on SrTiO$_3$ heterostructures. 
Owing to the broken inversion symmetry at the interface/surface, Cooper pairs acquire a finite 
center-of-mass momentum in a magnetic field parallel to the two-dimensional conducting plane.
It has been demonstrated that unconventional FFLO states emerge in the high-carrier-density region 
owing to the orbital degree of freedom in $t_{2g}$ electrons.

For $\HH \parallel$ [100], the Cooper pair momentum shows a nonmonotonic magnetic field dependence 
indicating the crossover in the superconducting state. 
Indeed, the magnetic field changes the superconducting state from the two-dimensional helical state 
to the quasi-one-dimensional helical state, and a higher magnetic field stabilizes the quasi-one-dimensional FF state. 
Near the crossover from the helical state to the FF state, the Cooper pair momentum discontinuously changes, 
and the double-$Q$ CS state is stabilized below $T_{\rm c}$. 

For $\HH \parallel$ [110], the double-$Q$ ODCS state is stabilized in the high-magnetic-field region. 
Therein, Cooper pairs formed by the d$_{yz}$-orbital and those formed by the d$_{zx}$-orbital have inequivalent 
center-of-mass momenta. 
These behaviors have not been shown in single-band models, and they are indeed attributed to 
the orbital degeneracy between the (d$_{yz}$, d$_{zx}$)-orbitals. 
Thus, the ODCS state is a novel FFLO state in the multi-orbital system.

The unusual properties of the FFLO state disappear in the low-carrier-density region where the 
superconductivity is mainly induced by the single d$_{xy}$-orbital. Thus, a high carrier density 
is needed to stabilize the multi-orbital FFLO state studied in this paper. 
Combining our results with the experimental observation,~\cite{joshua2012universal} 
we expect that SrTiO$_3$ heterostructures are in the high-carrier-density region when $n > 2 \times 10^{13}$ cm$^{-2}$. 
Two-dimensional electron gases with high carrier densities beyond $n = 1 \times 10^{14}$ cm$^{-2}$ have already been 
realized.~\cite{Ueno,Ueno_review}

Experimental studies of the superconducting state in a high magnetic field or in the ferromagnetic state 
are desired to clarify the FFLO state in SrTiO$_3$ heterostructures. 
We showed that large in-plane anisotropy in the critical magnetic field will be evidence 
of the multi-orbital superconducting state. 
More direct observation of the FFLO state using scanning tunneling microscope (STM) and other techniques is of course desired. 
We hope that our proposal will be verified.

Finally, we discuss two ingredients that are not taken into account in our model. 
One is the subband structure. According to the band structure calculations,~\cite{
delugas2011spontaneous,pentcheva2008ionic,popovic2008origin,khalsa2012theory,hirayama2012ab,zhong2013theory,King2013} 
several d$_{xy}$-orbital-derived subbands cross the Fermi level. On the other hand, the d$_{yz}$/d$_{zx}$-orbitals are 
quantized in the [001]-axis direction because of their light effective mass. 
The subband structure in the d$_{yz}$/d$_{zx}$-orbitals is negligible, and thus the single-layer model adopted in this work 
is justified. 
Although the d$_{xy}$-orbital-derived subbands reduce the carrier density in the d$_{yz}$/d$_{zx}$-orbital-derived band, 
our results in the high-carrier-density region are not altered as long as a substantial part of carriers has a 
d$_{yz}$/d$_{zx}$-orbital property. 
The other ingredient neglected so far is the disorder. 
Although we investigated the superconducting state in the clean limit, the SrTiO$_3$ heterostructures 
indeed contain substantial disorder.~\cite{Bell,Ueno} The disorder may alter our results for the multi-orbital FFLO state 
because it smears out the orbital character of the band. Furthermore, it has been shown that the disorder 
destabilizes the FFLO state,~\cite{FFLO_review} although the helical state is not completely eliminated~\cite{NCSC}. 
Therefore, it is desirable to take into account the disorder for a comparison with experimental results. 
We leave this issue for a future study. It is also desirable to fabricate (at least locally) 
clean SrTiO$_3$ heterostructures.

\section*{Acknowledgements}

The authors are grateful to T. Nojima and K. Ueno for fruitful discussions. 
This work was supported by a ``Topological Quantum Phenomena'' (No. 25103711) 
Grant-in Aid for Scientific Research on Innovative Areas from MEXT, Japan, 
and by a JSPS KAKENHI (No. 24740230). 
Y. N. was supported by a JSPS Fellowship for Young Scientists.
Part of the numerical computation in this work was carried out 
at the Yukawa Institute Computer Facility.

\bibliographystyle{jpsj}
\bibliography{66799}

\end{document}